\documentclass[a4paper]{jpconf}
\usepackage{graphicx}
\begin{document}
\newcommand{\beq}{\begin{equation}}
\newcommand{\eeq}{\end{equation}}
\newcommand{\beqa}{\begin{eqnarray}}
\newcommand{\eeqa}{\end{eqnarray}}

\title{Antimatter and Gamma-rays from Dark Matter Annihilation}

\author{Lars Bergstr\"om}

\address{Department of Physics, Stockholm University\\
 AlbaNova University Centre, SE-10691 Stockholm, Sweden}

\ead{lbe@physto.se}

\begin{abstract}
A brief review of the indirect detection signatures of dark matter is given. In particular, detection
methods of dark matter particle annihilation to antimatter and $\gamma$-rays are reviewed.
With the GLAST satellite soon to be launched, a crucial window in the 
energy range of a few GeV up to 300 GeV will open. The good angular and energy
resolution of the instrument means that structures predicted by cold dark 
matter  models can be searched for. Large, currently planned  ground-based imaging Cherenkov telescope arrays, may further improve the limits, or discover a
signal, if the current understanding of halo dark matter structure is
correct.  
\end{abstract}



\section{Introduction}

The dark matter problem is one of the most obvious unsolved problems
of present-day cosmology. A wealth of data support the existence 
of non-baryonic dark matter, contributing on the order of a quarter
of the energy density of the Universe \cite{scopel}.
A particularly interesting class of dark matter models, with basis
in particle physics are the
so-called WIMPs - weakly interacting massive particles. The idea that
something new related to electroweak symmetry breaking could play a 
role for dark matter is a particularly interesting one, since it
gives a relic density of the required order of magnitude, and it
may with its electroweak interactions be detectable.

\section{Methods of detection}

Essentially, there  are three methods of detecting WIMPs. First,
indications of their existence may be obtained at accelerators,
in particular at CERN's upcoming LHC, where in favourable cases also the  
mass and some couplings to other particles could be measured.
However, in most cases it will be difficult to verify that 
new particles discovered at LHC are indeed the dark matter.
Thus, detection by some other methods may be crucial for 
identification. This problem has recently been studied in some
detail for supersymmetric models \cite{baltz}. Although supersymmetric
models are by no means the only contenders for dark matter they are 
so well motivated by other reasons (such as, e.g., solving the hierarchy
problem of particle physics, i.e., to explain why the mass hierarchy 
$m_{W}<<m_{Pl}$ is stable against radiative corrections), that 
the lightest supersymmetric particle, most probably the lightest
neutralino, has become something of a template for dark matter. One
of the prime objectives of LHC is to search for supersymmetry and
if found to determine as much as possible about its parameters.

The second way to search for dark matter employs sensitive detectors,
typically
consisting of sizable amounts of Na I or Ge crystal \cite{belli,cdms} or liquid
noble gas \cite{xenon}. After many years of steady but slow progress,
the direct detection technique now delivers quite impressive data, with
experiments now putting limits deep into the supersymmetric parameter
plane (for a since several years claimed, but unconfirmed, detection, 
see \cite{belli}). The present generation of detectors, of $50-100$
kg of detector material, is presently being enlarged towards the 1000 kg 
mass scale, which will test WIMPs models, including supersymmetric ones,
quite strongly.

Both accelerator searches and direct detection experiments may give 
indications of the properties of the particle making up the dark matter
(given that it is a WIMP, which is the scenario we treat here). However,
the third method, indirect detection of the final states produced in
dark matter particle annihilation, would in some sense be the most convincing
way to determine what makes up dark matter halos, like the one where the
Milky Way is embedded. The types of final state particles that may be detected
are antimatter (positrons and antiprotons, which are usually a very small
part of the ordinary cosmic rays), neutrinos and $\gamma$-rays. For supersymmetric
dark matter, one of the theoretically favoured templates of a WIMP, there
exist extensive computer packages like DarkSUSY \cite{darksusy}
or Micromegas \cite{micromegas} which allows to compute expected detection
rates for most of the processes of direct and indirect detection 
discussed here.

\section{Neutrinos}

Fluxes of neutrinos that are in principle measurable can be produced by 
annihilations of WIMPs trapped gravitationally at the center of the Earth
or the Sun \cite{sun_earth}. The mechanism of trapping, however, involves scattering on
nuclei which means that the capture rate and therefore in most cases
the annihilation rate from the Earth is limited by direct detection bounds. 
These rates have to be equal in steady state; 
an exception
may occur if the scattering rate is so small that the distribution has
not yet equilibrated. Usually one then has unobservable neutrino rates,
so generally it is the elastic scattering cross section, not the
annihilation cross section, that determines the annihilation rate \cite{JKG}.
This means that the neutrino rate from the Earth is most likely too small
to be detected in presently built neutrino telescopes (IceCUBE and Antares),
given the present rapidly improving bounds on the spin-independent
scattering cross sections \cite{cdms,xenon}, which are dominant given the
chemical composition of the Earth.

The Sun, however, contains mostly protons, so there the spin-dependent part of 
scattering is more important. The spin-dependent scattering cross section 
is notoriously difficult to bound experimentally, so therefore a possibility
for a signal from the interior of the Sun remains. In fact, even if the 
bounds on spin-dependent scattering would improve significantly, the 
spin-dependent scattering rate is so poorly correlated with annihilation
rate that the possibility will still exist \cite{beg_nu}. The rate predicted
for minimal supersymmetric models are reasonably high. For Kaluza-Klein
dark matter particles in universal extra dimensions the predicited rates
are, however, probably too small to be measurable at IceCUBE \cite{hooper_kribs}. 
\section{Antimatter}

\subsection{Positrons} 
Positrons  usually do not give
a very competitive method of discovery, as very large factors, ``boost factors''
are needed \cite{th_positrons} to contribute considerably to the 
measured positron spectrum
\cite{exp_positrons}.
This boost factor is an arbitrary factor used to enhanced the rate of
positrons over the flux one would get using standard halo models and
propagation properties. It could, for instance, be caused by small-scale
clumping of the halo, although most estimates would tend to maximize this
to $f_B\sim 2 -8$ \cite{venya}, perhaps being possible to push by another order
of magnitude at most \cite{diemand}. 
The other argument against using positrons
as a tools for dark matter detection is that they are easily produced
in large quantities by all acceleration processes, such as in supernova winds or
black hole accretion. Any spectral feature is likewise mostly erased due
to the large and rapid energy losses of positrons propagating through the 
Galaxy. An exception may be for very massive particles annihilating directly 
into
an electron-positron pair, such as happens in Kaluza-Klein models with universal
extra dimensions \cite{hooper}. Also there large boost factors are
needed, however. 

In priciple, large boost factors may be obtained if there happens to be 
a clump of dark matter near the solar system \cite{cumberbatch}. 
The probability for this to 
happen is, however, extremely small for realistic models of the distribution
of substructure in the galaxy \cite{salati_mm}.

\subsection{Antiprotons}
Antiprotons would give a more robust detection rate, although a rather
featureless modification of the spectrum from what is expected by cosmic-ray
induced antiprotons \cite{antiprotons}. 

The propagation of antiprotons is similar to that of positrons, a main difference being that the synchrotron
and inverse Compton energy losses are not important. Instead losses caused 
by elastic and inelastic scattering
on gas and dust in the galaxy become more important. This can be relatively 
well modeled as the amount of
material encountered during an energy loss time can be estimated, 
and the propagation parameters can be tuned
to give a ratio of secondary to primary nuclei (like $B/C$) that agrees 
with the observed value.

Early on it was thought that a unique feature of antiprotons produced in 
dark matter annihilations would be that they populate all of the available 
phase space down to very low energies as seemed to be indicated by
experimental data, whereas the background produced
by antiproton production by ordinary cosmic ray protons hitting hydrogen 
or helium would for kinematical reasons populate higher energies only. However, 
it was later shown that several effects, in particular energy loss
by elastic scattering of antiprotons on nucleons, tend to populate the 
low-energy part of the spectrum
 \cite{BEU_antiprotons}, 
giving a spectrum in excellent agreement with the measured one. 

Antiprotons probe a larger part of the galaxy surrounding the observation 
point than does positrons (for instance,
roughly half of the antiprotons produced near the galactic 
center make it out to the solar radius).  Their 
diffusion properties are also well known, which means that the 
agreement between the observed flux and the
cosmic-ray induced flux can be used to limit "exotic" contributions 
from dark matter annihilation in models
with large boost factors. One example is the model of de Boer \cite{deboer} 
which claims to fit well the $\gamma$-ray
flux measured by the {\sc Egret} experiment, by introducing a substantial 
contribution from dark matter 
annihilation. This is done by having a rather unusual distribution of dark 
matter (disk-concentrated with rings)  in the galaxy, and in addition working
with large boost factors. It seems that this model taken at face value would 
produce an antiproton 
flux which is an order of magnitude
too large to agree with data \cite{beg_salati}.

It is possible, though, that high-mass models of dark matter could give a 
measurable disturbance on the 
antiproton distribution at higher energies, something which will be 
interesting to see {\sc Pamela} \cite{pamela} 
investigate. 

\subsection{Antinuclei}
It is very unlikely that other antinuclei other than antiprotons are 
produced by cosmic ray interactions.
Therefore, even a handful of antideuterons would point to some exotic source. 
It was suggested
\cite{donato} that neutralino annihilations could be such a source of 
antideuterons, produced by the
accidental overlap of the wave functions of an antiproton and an antineutron 
produced at low relative velocity
in the quark fragmentation process. This idea is very interesting, but the 
uncertainties in the
theoretical estimates of the formation rate are hard to estimate 
(for an attempt, see \cite{baer}). 

\section{Gamma-rays}

We finally come to one of the most actively pursued indirect detection methods,
that of $\gamma$-rays created in WIMP annihilations \cite{sciama}. It is especially important
for the new large imaging air Cherenkov telescopes currently operating
(e.g., VERITAS, HESS  or CANGAROO \cite{hessetal})
or being planned \cite{agis}, and for the GLAST ($\gamma$-ray large area space telescope)
satellite planned to be launched in 2008 \cite{glast}. 
There are several recent reviews in this field (\cite{BUB,reviews}) - here we
only point out some of the recent highlights.

The annihilation rate towards a direction making the angle $\Psi$ with respect
to the galactic centre is conveniently given by the factorized expression
\cite{BUB}
\beqa
\Phi_{\gamma}(\psi) & \simeq & 0.94 \cdot 10^{-13}\left( \frac{N_{\gamma}\;v\sigma}
{10^{-29}\ {\rm cm}^3 {\rm s}^{-1}}\right)\left( \frac{100\,\rm{GeV}}
{M_\chi}\right)^2 \cdot \nonumber \\
&&\;\;\;\;\;\;\, J\left(\psi\right)\;\rm{cm}^{-2}\;\rm{s}^{-1}\;\rm{sr}^{-1}\label{eq1}
\eeqa
where we have defined the dimensionless function 
\beq
J\left(\psi\right) = \frac{1} {8.5\, \rm{kpc}} 
\cdot \left(\frac{1}{0.3\,{\rm GeV}/{\rm cm}^3}\right)^2
\int_{line\;of\;sight}\rho^2(l)\; d\,l(\psi)\;,
\label{eq:jpsi}
\eeq
with $\rho(l)$ being the dark matter density along the line of sight $l(\Psi)$.
(Note the numerical factor in Eq.~(\ref{eq1}) differs by a factor $1/2$ from that given
in \cite{BUB}; this takes into account the fact that the annihilating particles
are identical, as is the case for supersymmetric neutralinos. See the 
footnote in connection to Eq.~(21) of the publication \cite{beul} for a detailed explanation.)

The particle physics factor $N_\gamma v\sigma$, which is the annihilation rate
times the number of photons created per annihilation, can usually be rather accurately
computed for a given dark matter candidate. DarkSUSY uses PYTHIA \cite{pythia}
to estimate the number of photons with continuum energy distribution created by
quark fragmentation for given standard model final states. Special calculations
are needed for higher order QED corrections, which may be very important as we will see.

There have been many improvements of the treatment that was first performed 
in \cite{BUB}, and many interesting new promising features keep 
appearing, of which can be mentioned:
\begin{itemize}
\item{The existence of subhalos within dark matter halos \cite{venya}}.
As N-body simulations have become more refined (see, e.g., \cite{nbody}
for a recent example) it seems clear that the hierarchical formation
of smaller structures that merge into larger ones will leave intact 
a population of subhalos which are not destroyed as far as they can 
be followed by the simulations. As the smaller subhalos  formed
at an earlier epoch, when the average density of the universe was larger,
this may boost annihilation rates by a rather significant factor \cite{nbody}.
\item{The possible existence of intermediate mass black holes.}
The black hole at the galactic centre may cause an enhancement of the dark 
matter density around it (a ``spike'') \cite{gondolo_silk}, although 
this is not mandatory, and may depend sensitively on the merger history
of the black hole \cite{ullio_zhao_kam}. If there would exist a 
population of intermediate
mass black holes, the boosting of density in their vicinity may be operative
however, and may give observational $\gamma$-ray annihilation rates to be searched
for by GLAST \cite{silk_bertone_zen}. 
These objects would appear as point sources without optical
counterpart, and may be studied further by air Cherenkov telescopes once their
location has been established by GLAST. In fact, a few of the very 
recently discovered very faint dwarf galaxies may give detectable signals
if their dominant mass is indeed composed of WIMPs \cite{koushiappas}.

\item{The presence of dark matter streams \cite{sikivie,newberg,helmi}.}
The microscopic structure of the dark matter halo may also contain dark
matter streams, created by late infall of dark matter particles on an
existing halo (for an early, simplified model, see \cite{sikivie}), or by tidal
stripping of dark matter particles from dwarf galaxies \cite{newberg}.
N-body simulations are presently not accurate enough to follow these
details, that generally appear below the resolution limit. It can be argued,
however, that especially for direct dark matter detection, and indirect
detection through neutrinos and $\gamma$-rays, which all depend sensitively
on the dark matter density at specific positions in the halo, these effects
can be very important. Recently, a new interesting method to compute these and similar
effects has been proposed \cite{helmi}.  

\item{The possibility of an extragalactic signal \cite{beu_extragal}.}
The fact that the combined structure formation in the universe would 
cause an increase of the extragalactic rate (i.e., integrating along the 
line of sight back to arbitrary large redshift, and combining with the
estimated absorption of $\gamma$-rays an optical and IR photons), by a factor
of order $10^5$ to $10^6$, was only recently realized 
\cite{beu_extragal,taylor_silk,beul,mannheim}. This is an area where
observations by GLAST will be crucial.

\item{``Explosive'' annihilation.} This phenomenon, discovered in \cite{explosive}, and verified in \cite{explo2}  gives the possibility of very strong 
$\gamma$-ray signals 
for particular masses (usually in the TeV region). Of course, TeV particles
may have difficulty to give the required relic density, unless one tolerates
some fine-tuning, as is explicitly done in ``split SUSY'' models \cite{giudice}.
\item{The discovery of models where the $2\gamma$ and $Z\gamma$ lines would actually be the dominant features of the $\gamma$-ray spectrum \cite{idm}.} These
are ``inert'' Higgs models with an extra Higgs doublet protected by a $Z_2$
symmetry. Besides other intersting phenomelogy, these models may also break
electroweak symmetry radiatively \cite{tytgat}.
\item{Large contributions at high energy from internal QED bremsstrahlung.}
 This interesting
phenomenon, originally proposed in \cite{lbe89}, which one has to go beyond 
leading order to see, has 
recently been proposed as a method to detect an otherwise undetectable
dark matter candidate \cite{baltz_bergstrom}. It has also been applied to
MSSM and mSUGRA models, and found to be very important \cite{torsten}, causing
sometimes large boosts to the highest energy end of the $\gamma$-ray spectrum.
The idea, for fermion final states, is that a 
Majorana fermion (as many dark matter candidates are) suffers a helicity
suppression for S-wave annihilation \cite{goldberg}. However, by emitting
a photon from an internal charged leg, which only costs a factor of
$\alpha_{em}/\pi$, the helicity suppression may be avoided. The effect will
be that these radiative corrections, instead of as usual being a percent
of the lowest order process, may instead give enhancement {\em factors}
of several thousands to millions times the suppressed lowest order,
low-velocity, rate \cite{lbe89}.
The resulting spectra will have a characteristic very sharp drop at
the endpoint $E_\gamma=m_\chi$ of the $\gamma$-ray spectrum, see Fig.~1.

\end{itemize}

\begin{figure}
\begin{center}
\includegraphics[width=80mm]{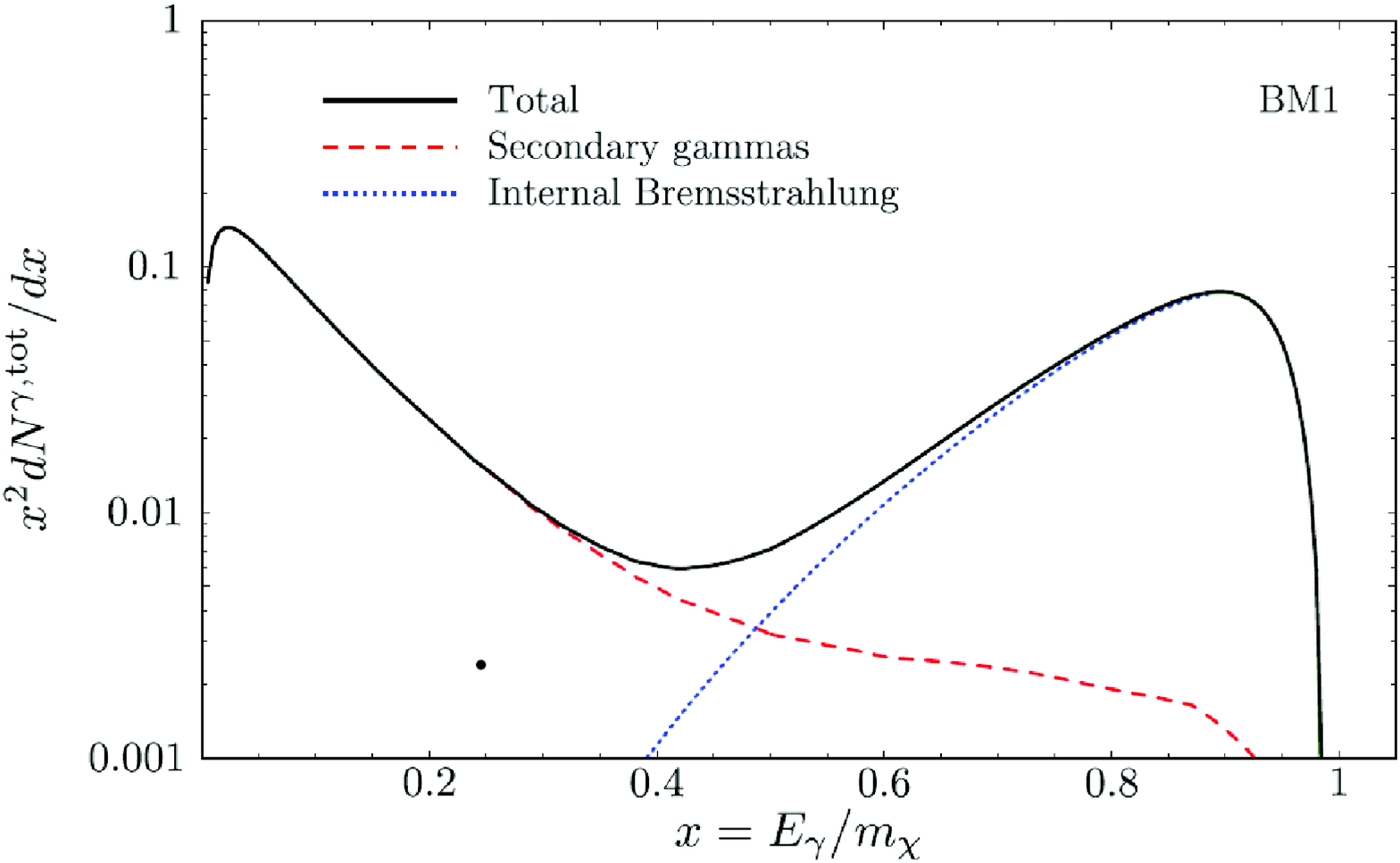}
\end{center}
\caption{\label{label}The $\gamma$-ray spectrum expected for one of 
the benchmark models (BM1) defined in \protect\cite{torsten}. Here $x=E_\gamma/m_\chi$,
and the distribution without the bremsstrahlung is shown by the dashed line.
There should also be a $2\gamma$ and a $Z\gamma$ contribution, not shown here.}
\end{figure}

\section{Connection between antimatter and $\gamma$-ray signal}

Recently, W. de Boer and colleagues have argued that the spectral distortion,
the so-called ``GeV anomaly'' of EGRET's measurements of galactic diffuse
$\gamma$-rays 
can be explained by supersymmetric WIMP annihilation \cite{deboer}. 
As explained earlier, this is done by 
making a very particular model of the distribution of dark matter,
which is largely concentrated to the disk, and
in addition has two very massive rings of dark matter in the plane
of the disk. It is claimed that by using this dark matter distribution,
and adjusting the annihilation rate by a rather large ``boost factor'',
the measured $\gamma$-ray angular and energy distributions can be reproduced
with a canonical supersymmetric WIMP, of mass below 100 GeV.

It is actually still not known whether the GeV excess of EGRET is real, 
or just an 
(unknown) experimental artifact as has been recently claimed by some members of
the collaboration \cite{stecker}. In any case, we mentioned that the proposal of de Boer et al.
has the problem of a severe overproduction of antiprotons \cite{antipr}. 
Given the
already somewhat contrived nature of the dark matter distribution, it is
difficult to consider very seriously attempts to change the diffusion
properties of antiprotons, as maybe could be a way out \cite{deb}. We rather
note that the rates for continuum gamma rays and antiprotons are strongly correlated for dark matter candidates that annihilate into standard model particles
as they are both given by the fragmentation of quark jets, 
and are reasonably well 
understood after a decade of LEP observations. On the other hand, 
this also means
that if one restricts oneself to standard propagation models, and dark matter models which fulfill the constraints from
present antiproton measurements (we of course soon anticipate new
interesting data from the PAMELA satellite \cite{pamela} which has been in orbit for well over a year), indications are that the detection of a $\gamma$-ray
signal has to await GLAST, and possibly the next generation of imaging air Cherenkov
telescope arrays. 
\subsection*{Acknowledgement}
This work was supported by the Swedish Research Council (VR).

\smallskip
\end{document}